\documentclass[article,preprint,floatfix]{revtex4-1}
\usepackage{textcomp}
\usepackage{graphicx}
\usepackage{amssymb}
\usepackage{epsfig}
\usepackage{ifsym}
\usepackage{wasysym}
\usepackage{latexsym}

\begin{document}

\title{Electron transport of WS$_2$ transistors in a hexagonal boron nitride dielectric environment}

\author{Freddie Withers, Thomas Hardisty Bointon, David Christopher Hudson, Monica Felicia Craciun and Saverio Russo}
\address{Centre for Graphene Science, College of Engineering, Mathematics and Physical Sciences, University of Exeter, Exeter EX4 4QF, UK}

\begin{abstract}
We present the first study of the intrinsic electrical properties of WS$_2$ transistors fabricated with two different dielectric environments WS$_2$ on SiO$_2$ and WS$_2$ on h-BN/SiO$_2$, respectively. A comparative analysis of the electrical characteristics of multiple transistors fabricated from natural and synthetic WS$_2$ with various thicknesses from single- up to four-layers and over a wide temperature range from 300K down to 4.2 K shows that disorder intrinsic to WS$_2$ is currently the limiting factor of the electrical properties of this material. These results shed light on the role played by extrinsic factors such as charge traps in the oxide dielectric thought to be the cause for the commonly observed small values of charge carrier mobility in transition metal dichalcogenides. 
\end{abstract}

\maketitle

\textbf{\large{Introduction}}
\\
The emerging class of atomically thin semiconducting materials formed by transition metal dichalogenides (TMDCs) is showing a plethora of complementary properties to those of graphene that are of interest to fundamental and applied research. These materials are uniquely suited to study the superconducting phase transition in the extreme two-dimensional limit inherent to atomically thin systems \cite{Novoselov2005,Ayari2007,Taniguchi2012,Bao2013}. At the same time TMDCs have a band gap which is essential for transistor applications and which could enable a new class of atomically thin photo-transistors. For example WS$_2$ has a direct band gap of 2 eV in single layer form \cite{Band1,gap1,gap2,Gutierrez} and has already shown great promise as a flexible transistor with field effect mobilities comparable to the best liquid crystals and on/off ratio of the current exceeding 10$^6$ \cite{liquid}. Understanding the limiting factors of the electrical properties of TMDCs is an open quest and a stepping stone for accessing novel physics in these systems. 

The typical values of charge carrier mobility measured in thin WS$_2$ flakes are always much lower than those measured in bulk material \cite{Ayari2007,Bao2013}. This behaviour has been interpreted as due to defect states in the SiO$_2$ substrate leading to the localization of charge carriers in TMDCs and a small charge carrier mobility \cite{impurity}. To probe the intrinsic electrical properties of TMDCs it would be necessary to measure electrical transport in either suspended structures or in transistors fabricated on clean substrates with fewer impurities than typically present in SiO$_2$. An ideal choice for such a substrate is hexagonal boron nitride \cite{hBN}, which is a preferred substrate for high quality graphene transistors since it has a very low concentration of charge scattering impurities and is atomically flat \cite{highMob}. To date such a study has not yet been conducted and the consequent lack of knowledge is limiting the potential impact of TMDCs on fundamental and applied research. Furthermore most of the studies conducted so far have been limited to just MoS2, while other TMDCs such as WS2 have not yet received much attention, whereas they might be better suited than MoS$_2$ for a given application. 

Here we present the first study of the electrical properties in WS$_2$ transistors fabricated on different dielectrics (i.e. SiO$_2$ and h-BN/SiO$_2$) and using synthetic as well as natural WS$_2$. The comparative analysis of the electrical characteristics of these transistors studied in the temperature range from 300K down to 4.2 K shows that in all cases electrical transport takes place \textit{via} hopping conduction through localized states \cite{shklov,Efros}. At low temperature (T$<$20K) we observe peaks of the conductance as a function of back-gate voltage and source-drain bias due to inelastic tunnelling in the impurity states with sub-gap energy. These results show that intrinsic disorder rather than extrinsic factors such as defect states in the oxide dielectric is limiting the electrical properties of WS$_2$ and more generally TMDCs. 

\textbf{\large{Results}}
\\
Thin flakes of WS$_2$ were obtained by mechanical exfoliation of flakes from synthetic crystals onto p-doped Si/SiO$_2$ substrate that serves as a back gate (for natural WS$_2$ see supporting information). Thin flakes are first identified with the aid of optical microscopy and their thickness is subsequently determined by atomic force microscopy (AFM) and Raman spectroscopy. The fabrication of WS$_2$ transistors on h-BN and subsequent encapsulation in h-BN is carried out using the dry transfer method first developed for graphene \cite{highMob}. This consists of exfoliating WS$_2$ onto a substrate coated by water soluble polymer and PMMA. After dissolving in water the soluble polymer, the free WS$_2$/PMMA bilayer is aligned onto previously exfoliated h-BN ($\sim$ 20 nm thick) on p-doped Si/SiO$_2$. The substrate is then heated up to melt the PMMA and secure contact between WS$_2$ and h-BN and the PMMA is subsequently removed in acetone. Electrical contacts to WS$_2$ are fabricated using standard electron beam lithography, thermal evaporation and lift-off of Cr/Au (5/70 nm). 

\begin{figure}[ht]{}
\includegraphics[width=0.9\textwidth]{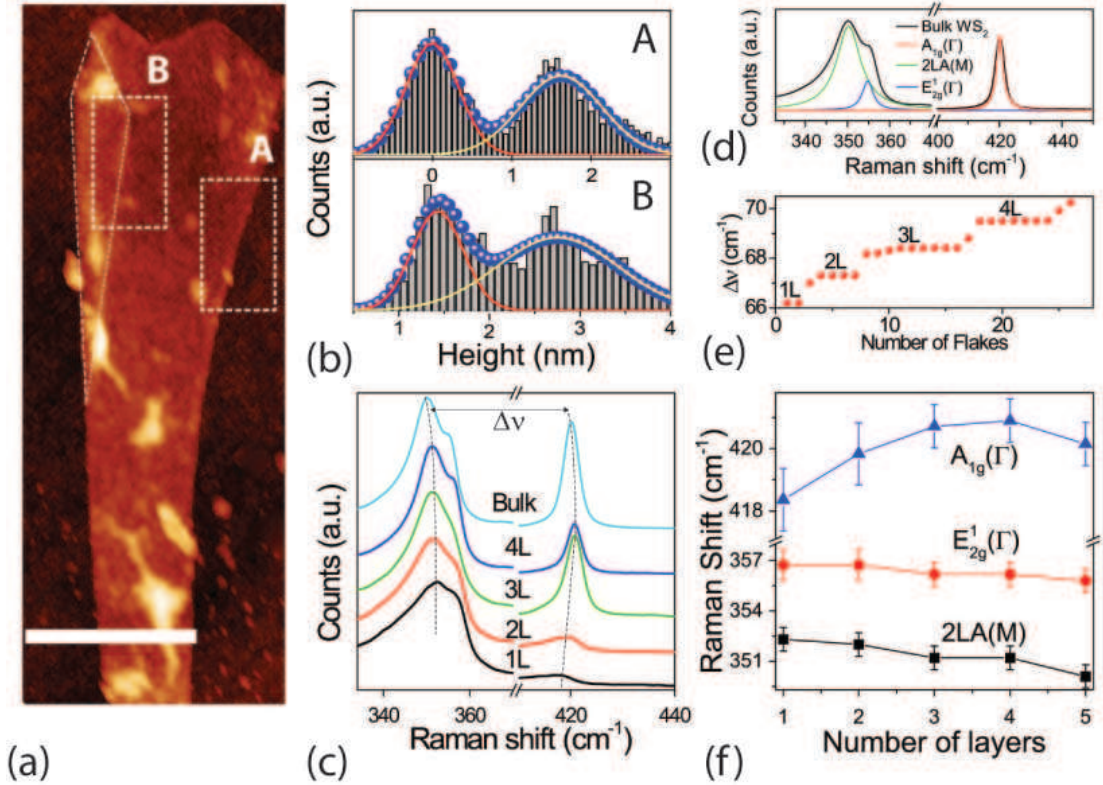}
\caption{\label{fig1} (a) shows an AFM measurement of a bilayer WS$_{2}$, the scale bar corresponds to 500 nm. The dashed areas labeled by A and B enclose the step edge at the SiO$_2$-bilayer WS$_2$ and the fold in the WS$_2$ flake respectively. The corresponding histograms of the measured heights in A and B are shown in (b). (c) shows the evolution of the shape and position of the Raman peaks of WS$_2$ as the number of layers is increased from single layer to bulk. (d) is a plot of the Raman spectra for bulk WS$_2$ and a fit to three Lorentzians corresponding to the 2LA(M), E$_{2g}^1(\Gamma)$ and A$_{1g}(\Gamma)$, see main text. (e) shows the measured wavenumber shift ($\Delta \nu$) between 2LA(M) and A$_{1g}$ plotted for 30 flakes with different layer number. (f) summarizes the measured Raman shift for 2LA(M), E$_{2g}^1(\Gamma)$ and A$_{1g}(\Gamma)$ as a function of layer number.}
\end{figure}

Figure 1a shows an AFM measurement of a thin WS$_2$ flake with a fold in the upper left corner highlighted by a dashed line. A statistical study of the height measured in areas which include the step edge at WS$_2$/SiO2 (region A) and the folded corner (region B) shows a comparable step height of $\approx 1.6$ nm in A and $\approx 1.3$ nm in B, see Figure 1b. Since the thickness of a monolayer WS$_2$ flake is $\approx 0.65$ nm \cite{Ramanws3, Gutierrez} we conclude that this flake is a bilayer. A comparative plot of the Raman spectra (see methods) for WS$_2$ with different layer numbers shows marked differences depending on the specific thickness of the flake, see Figure 1c. More specifically it is known that the peak with low Raman shift ($\approx 350$ cm$^{-1}$) is a convolution of two Lorentzians (Figure 1d) whose positions change as a function of the layer number \cite{Ramanws2}. One Lorentzian is due to the second order longitudinal acoustic phonon mode (2LA(M)) corresponding to collective oscillations of the atoms in the plane, and this gives a Raman peak at 352.7 cm$^{-1}$ in single layer WS$_2$. The second Lorentzian is given by the in-plane optical phonon mode (E$^1_{2g}$($\Gamma$)) representing the in-plane counter oscillations of W and S atoms in the lattice. Finally the out-of-plane optical phonon mode (A$_{1g}$($\Gamma$)) representing the out-of-plane oscillations of W and S atoms gives a Raman peak at 416.6cm$^{-1}$. A plot of the relative wavenumber shift ($\Delta\nu$ that is the difference between the 2LA(M) and A$_{1g}$($\Gamma$)Raman peaks) for a large number of flakes with various thicknesses shows that $\Delta\nu$ changes in a discrete way according to the number of layers which have been independtly measured with AFM, see Fig. 1e. Finally, upon increasing the number of WS$_2$ layers the position of the 2LA(M) and E$^1_{2g}$($\Gamma$) peaks redshift monotonously, whereas the A$_{1g}$($\Gamma$) peak blue shifts as previously shown \cite{Ramanws2}, see Figure 1f.

\begin{figure}[ht]{}
\includegraphics[width=0.8\textwidth]{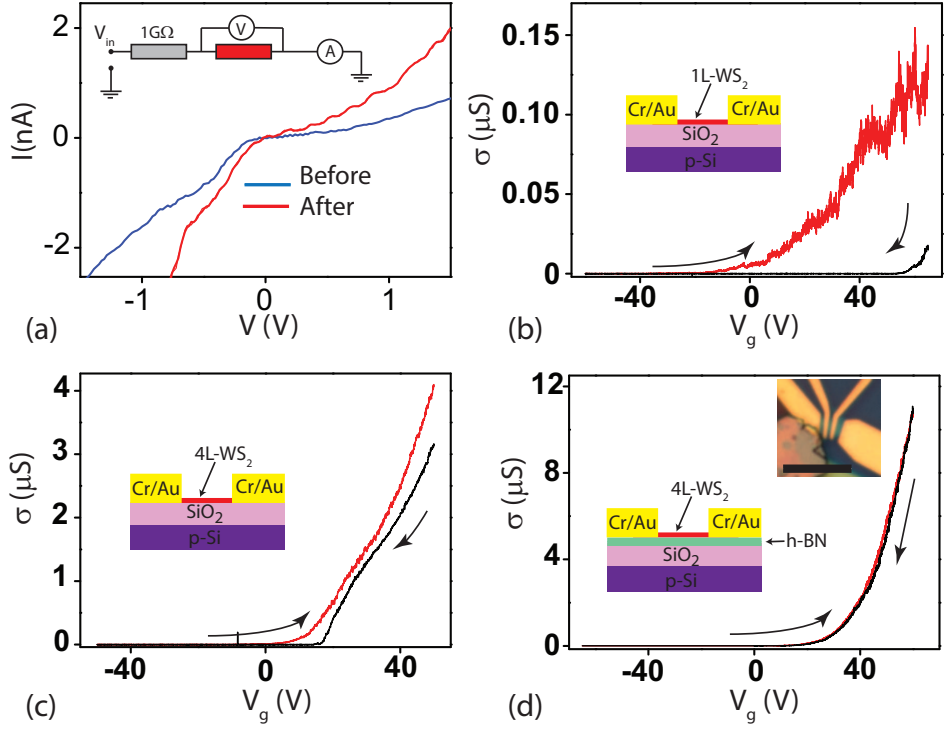}
\caption{\label{fig2} (a) shows plots of I-V for a 4L WS$_2$ device on a 20nm thick h-BN crystal before and after sweeoubg the voltage bias to large values (see supplementary information). The inset shows the circuit
used for decreasing the contact resistance. (b-d) are plots of the gate dependence of the conductivity for a monolayer WS$_2$ flake on SiO$_2$ substrate (b), for a four layer WS$_2$ flake on SiO$_2$ substrate (c) and for a 4 layer WS$_2$ flake on h-BN substrate (d). All the sweeps in (b-d) were made at the same rate of 100 V/hr. The inset in (d) shows a micrograph picture of a WS$_2$ transistor with scale bar of 5um.}
\end{figure}

Having established a reliable procedure to identify the layer number of WS$_2$ flakes we now turn to investigate the electrical transport properties of this material. The source-drain current \textit{vs.} bias voltage characteristics (I-V) of WS$_2$ transistor devices are always highly non-linear and upon performing current-bias annealing, a linear I-V around zero voltage bias is attained (see Figure 2a and supplementary information). Owing to the difference in work function between WS$_2$ and Cr, a Schottky barrier of about 100meV has to be expected at this interface when no-gate voltage is applied. The observed bias-annealing changes in the I-V and the large values of voltage bias at which these non-linearity occur suggest a different origin for this phenomenon, that is the possible presence of an oxide barrier at the WS$_2$/Cr interface which can be electrically broken upon applying a large voltage bias as shown in Figure 2a. In the following we only consider the analysis of electrical transport measurements in devices after bias-annealing.

Figure 2b-d show the room temperature field effect transistor (FET) transfer characteristics, that is the gate voltage (V$_g$) dependence of the conductivity ($\sigma$), for monolayer WS$_2$ on SiO$_2$ (Figure2b), four-layer WS$_2$ sample on a SiO$_2$ (Figure2c) and four-layer WS$_2$ sample on a  h-BN/SiO$_2$ (Figure2d). In all cases we observe that the conductivity has a large on-off ratio typical of semiconducting materials, with a finite threshold voltage. However we find that the field effect mobility ($\mu$) is always larger in WS$_2$ on h-BN than in WS$_2$ on SiO$_2$ (0.23 cm$^{2}$V$^{-1}$s$^{-1}$ for 1L-WS$_2$/SiO$_2$, 17 cm$^{2}$V$^{-1}$s$^{-1}$ for 4L-WS$_2$/SiO$_2$ and $\approx$ 80 cm$^{2}$V$^{-1}$s$^{-1}$ for 4L-WS$_2$/h-BN/SiO$_2$ in Figure 2(b-d)). A large hysteresis is also present in $\sigma(V_g)$ for WS$_2$ on SiO$_2$ but is fully suppressed when WS$_2$ is on h-BN/SiO$_2$. Similar hysteresis in I-V have also been reported in graphene and is commonly attributed to dopants present in the SiO$_2$ dielectric \cite{SiO2surface,grapheneHyst}.

\begin{figure}[ht]{}
\includegraphics[width=0.8\textwidth]{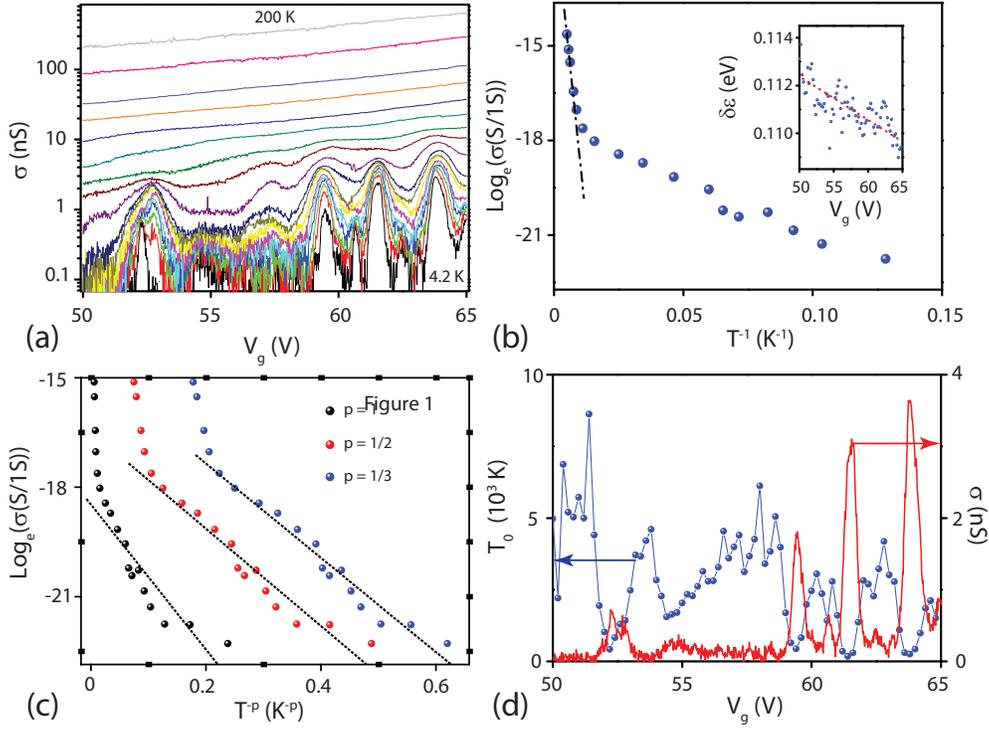}
\caption{\label{fig3} (a) shows a representative plot of $\sigma(V_g)$ for a 4L-WS$_2$ in the gate range 50 V to 65 V. (b) Typical temperature dependence of the conductivity plotted in terms of the activation energy relation (this curve is taken at V$_g$ = 60.5 V). The inset is a plot of the extracted activation energy for different gate voltages, each point corresponds to an average over 0.2 V gate voltage. (c) shows the same data as in (b) but plotted in terms of 2D Mott variable range hopping relation. (d) The conductivity at 4.2 K plotted alongside the hopping parameter T$_0$. A strong correlation between the two is observed: i.e. peaks in conductance correspond to low values of T$_0$.}
\end{figure}

\textbf{\large{Discussion}}

For all the measured devices we find that the temperature dependence of $\sigma(V_g)$ shows a pronounced suppression of the value of $\sigma$ upon lowering the temperature as expected for a semiconducting material, see Figure 3a. In these devices we apply a large enough value of gate voltage such that the charge carriers are directly injected from the metal contacts into the conduction band of WS$_2$. In this limit the relevant energy scale dominating the temperature dependence of the zero-bias resistance is the difference between the Fermi energy and the conduction band edge of the n-doped semiconductor (i.e. WS$_2$) \cite{Das2013}. A plot of $\sigma$ as a function of T$^{-1}$ at $V_g=60.5V$ reveals that from 260K down to 100K the conduction takes place by thermally activated charge carriers, i.e.  $\sigma(T)=\sigma_0 \ exp(-\delta \varepsilon/2 k_B T)$ with $\delta \varepsilon$ the activation energy and $k_B$ the Boltzman constant. The values of $\delta \varepsilon$ estimated from a fit of $\sigma(T)$ for 50V$<V_g<60$V are in the range 0.109eV$<\delta\varepsilon<$0.113eV and change linearly with $V_g$, see inset in Figure 3b. These values of $\delta \varepsilon$ are compatible with the voltage bias range over which non-linear I-V are measured (see blue curve in Figure 2a) suggesting that $\delta \varepsilon$ is the energy from the Fermi level to the conduction band edge ($E_c$), i.e. $\delta \varepsilon = E_c-E_F$ which is also much larger than the Schottky barrier height ($\approx$ 100 meV). 

The smooth dependence of $\delta \varepsilon $ on $V_g$ demonstrates that for sub-gap energies the Fermi level can be continuously tuned by means of a gate voltage throughout the defect induced states. To estimate the density of defect states we consider the equivalent gate capacitance of these WS$_2$ transistors that is the series of the gate oxide capacitance ($C_{ox}$) and defect states capacitance ($C_t$), i.e. $- dE_F/dV_g = 1.5\times 10^{-4}e = e\frac{C_{ox}C_t}{C_{ox}+C_{t}}$ \cite{Ayari2007}. Knowing that the oxide capacitance per unit area is $C_{ox} = \frac{\epsilon_r \epsilon_0}{d} \approx 1.2 \times 10^{-4}$ Fm$^{-2}$ we find $C_t = 0.8 Fm^{-2} = q^{2} D(E)$, where q is the unit of charge and D(E) is the density of defect states which we estimate to be $3.12 \times 10^{37}$ J$^{-1}$m$^{-2}$ .

\begin{figure}[ht]{}
\includegraphics[width=0.6\textwidth]{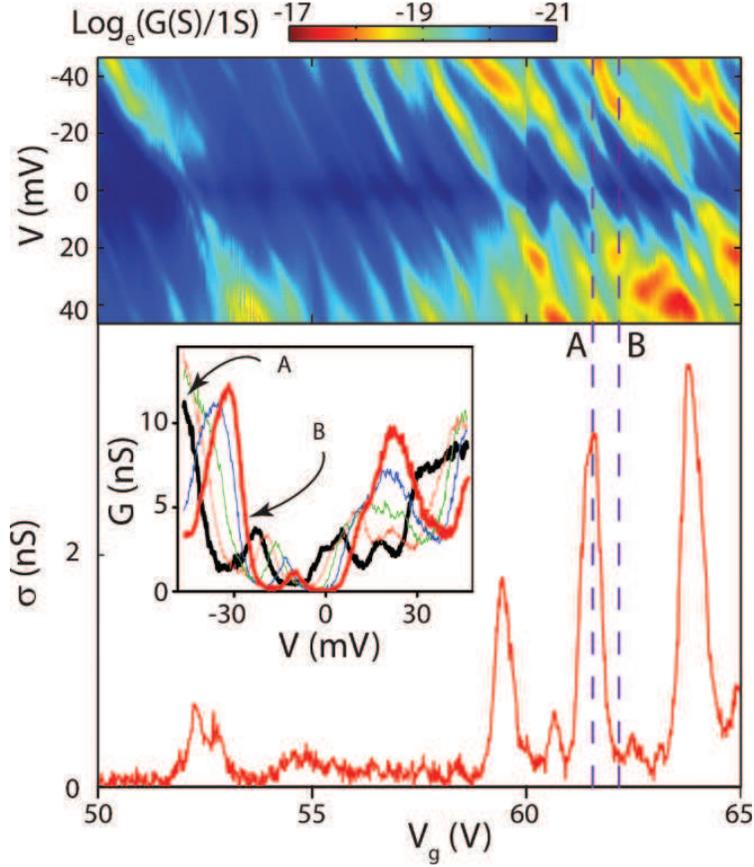}
\caption{\label{fig4} The top colour map shows the measured differential conductance plotted against gate voltage and source drain at T= 4.2K for the same representative 4L-WS$_2$ on h-BN discussed in Figure 3. The bottom plot is a graph of the conductivity at T= 4.2 K while the inset is a graph of five differential conductance curves plotted from V$_g$ = 61.52 V up to 62 V highlighted by the dashed lines A and B respectively and in steps of 96 mV of gate voltage.}
\end{figure}

The dominant role of disorder induced states with sub-gap energies becomes fully apparent when considering a fit of the low temperature $\sigma(T)$ in logharitmic scale in terms of $T^{-p}$ with $p$ critical exponent, see Figure 3c. This study reveals that p=1/3 gives the best fit stemming for non-interacting Mott variable range hopping \cite{shklov,Efros,Alex} $\sigma = \sigma_{0}exp(-T_{0}/T)^{1/3}$ where $T_{0}$ is the hopping parameter and is related to the density of localised states existing within the forbidden gap and the electron wavefunction size $\xi$ by the following relation $T_{0} = \frac{13.8}{k_B D_c \xi^2}$. The extracted values for the hopping parameter $T_0$ at each different gate voltage are plotted in figure 3(d) along with the conductance at T = 4.2 K. This comparative plot shows a clear correlation between the hopping parameter and the conductance whereby peaks in conductance correspond to very low values of $T_0$. Furthermore $T_0$ is found to fluctuate from $\approx 100$ K to $\approx 4000$ K in a small gate range (from $V_g=50.5V$ to 52V, corresponding to an energy window of just 0.25meV). Consequently the estimated localization radius in WS$_2$ increases from 1.8 nm to 17 nm. These observations indicate that the sub-gap impurities states have peaks of narrow energy band-widths dominating electrical transport for sub-gap energies.

Another prominent feature evident in the temperature dependence of $\sigma(V_g)$ is the emergence of peaks for T$<$100K with decreasing amplitude for T$<$20K, see Figure 3a. At the same time the differential conductance as a function of source-drain bias and gate voltage at T$=$4.2K (Figure 4a) shows that these peaks shift their position as a function of voltage bias. These observations suggest that charge transport at sub-gap energies occurs through inhomogeneous charge puddles and localized states in WS$_2$. Since we observe a similar $\sigma(V_g)$ behaviour in a variety of samples independently of (1) the WS$_2$ flakes aspect ratio, (2) the WS$_2$ layer number and (3) the dielectric environment (WS$_2$/BN/SiO2, see supplementary information) we conclude that the localized states dominating electrical transport in WS$_2$ at sub-gap energies are intrinsic to the WS$_2$ and not extrinsic such as defect states in the dielectric.  

To estimate the localization radius ($\xi$) we consider electrical transport measurements of a representative 4L-WS$_2$ in which the peaks of $\sigma(V_g)$ are spaced by an average gate voltage $\left\langle V_g \right\rangle \approx 1.13$ V corresponding to 0.17 meV, see bottom graph in Figure 4. In this device the peaks of $\sigma$ at fixed $V_g$ as a function of source-drain bias ($V$) are spaced by an average source-drain bias $\left\langle V \right\rangle \approx 11$ mV. Since the threshold voltage bias needed to observe electrical conduction is $\sim 11$ mV and the sample has a length of 350 nm the threshold electric field is $E_T = 3.14\times 10^4$ V/m. This value of $E_T$ together with the observed average peak separation of 0.17 meV gives a localization region of diameter $2 \xi=5.4$ nm which is consistent with the extracted value of the localization radius $\xi$ from the analysis conducted on the temperature dependence of $\sigma(Vg)$.

Finally we note that Coulomb blockade cannot account for the observed peaks of $\sigma(V_g)$. Indeed, if we assume a charging energy in our devices of Ec $\sim$ 40 - 50 meV estimated directly from the stability diagram shown in Figure 4, we extract a diameter d $\sim e^2/4\varepsilon_0 \varepsilon_r$Ec $\sim$ 20 - 40 nm for the confining regions ($e = 1.6 10^{-19}$C, $\varepsilon_0 = 8.85 10^{-12} F/m$ and $\varepsilon_r = (\varepsilon_{vac} +\varepsilon_{BN})/2= 2.5$ with the dielectric constant for vacuum and BN $\varepsilon_{vac}=1$ and $\varepsilon_{BN}=4$). Given the dimensions of the conductive WS$_2$ channel, our devices would consist of 100-1000 charging regions (i.e. (length $\times$ width)/d = (350 nm $\times$ 1500 nm)/d). The stability diagram of such an array of charging islands would consist of many overlapping Coulomb diamonds which are not observed in our measurements. An indication of the underlying physical process originating these peaks of $\sigma(V_g)$ is given by the temperature dependence of $\sigma(V_g)$ presented in Figure 3a: we alwasy observe that the amplitude of the peaks decreases upon lowering the temperature. This behaviour has been previously reported in other semiconducting systems \cite{Yakimov,Savchenko} and it is a fingerprint of inelastic tunnelling which in WS$_2$ occurs through the sub-gap impurity states.

In summary we have presented the first systematic study of the intrinsic electrical properties of thin WS$_2$ flakes. By comparing the I-V of transistors fabricated using two different dielectric environments (i.e. (1) WS$_2$ on SiO$_2$ and (2) WS$_2$ on h-BN/SiO$_2$) we find that hopping through localized states dominate electrical transport over a wide temperature range (T$<$100K). This intrinsic disorder has a finite density of states at sub-gap energies which contribute with inelastic tunnelling to electrical transport. These results demonstrate the dominant role played by intrinsic disorder over extrinsic factors such as defect states in the oxide dielectric as a limiting factor of the electrical properties of WS$_2$.   

\noindent \textbf{Methods}\\
Materials: Synthetic WS$_2$ was purchased from Lowerfriction.com.\\
Measurement techniques: The Raman spectra where measured with a Renishaw spectrometer using an excitation laser with a wavelength of 532 nm, focused to a spot size of 1.5 $\mu$m diameter and
1 mW incidente power. These measurements were performed in air and at room temperature. \\
Electrical measurements: The electrical transport measurements were performed in constant voltage configuration with excitation voltage smaller than k$_B$T, with k$_B$ Boltzmann constant. The differential conductance was measured using the lock-in technique.\\

\noindent \textbf{Acknowledgments}\\
F.W. acknowledges Gunnar F\"{a}rber (mineralien@online.de) for providing a specimen of natural tungstenite. 
S.R. and M.F.C. acknowledge financial support from EPSRC (Grant no. EP/J000396/1 and no. EP/K010050/1) and from the Royal Society Travel Exchange Grant 2012 and 2013.\\

\noindent \textbf{Author contribution statement}\\
F.W. conducted the fabrication and electrical measurements. T.H.B. conducted the AFM measurements. D.C.H. participated in the electrical measurements. F.W., S.R. and M.F.C. interpreted the data and wrote the manuscript. All authors reviewed the manuscript.

\noindent \textbf{Additional information}\\
Competing financial interests: The authors declare no competing financial interests.
License: This work is licensed under a Creative Commons
Attribution-NonCommercial-NoDerivs 3.0 Unported License. To view a copy of this
license, visit http://creativecommons.org/licenses/by-nc-nd/3.0/


\begin{thebibliography}{100}


\bibitem{Novoselov2005} Novoselov, K.S., \textit{et al.} Two-dimensional atomic crystals. \textit{P. Natl. Acad. Sci. USA} \textbf{102}, 10451-10453 (2005).

\bibitem{Ayari2007} Ayari, A., Cobas, E., Ogundadegbe O. $\&$ Fuhrer, M.S. Fabrication and electrical characterization of ultrathin crystals of layered transition-metal dichalcogenides. \textit{J. App. Phys.} \textbf{101}, 014507 (2007).

\bibitem{Taniguchi2012} Taniguchi, K., Matsumoto, A., Shimotani H. $\&$ Takagi, H. Electric-field-induced superconductivity at 9.4 K in a layered transition metal disulphide MoS$_2$. \textit{App. Phys. Lett.} \textbf{101}, 042603 (2012).

\bibitem{Bao2013} Bao, W., Cai, X., Kim, D., Sridhara, K. $\&$ Fuhrer, M.S. High mobility ambipolar MoS$_2$ field-effect transistors: Substrate and dielectric effects. \textit{App. Phys. Lett.} \textbf{102}, 042104 (2013).

\bibitem{Band1} Mattheiss, L.F. Band Structures of Transition-Metal-Dichalcogenide Layer Compounds. \textit{Phys. Rev. B} \textbf{8}, 3719-3740 (1973).

\bibitem{gap1} Frey, G.L., Elani, S., Homyonfer, M., Feldman, Y. $\&$ Tenne, R. Optical-absorption spectra of inorganic fullerenelike MS$_2$ (M=Mo, W). \textit{Phys. Rev. B} \textbf{57}, 6666-6671 (1998).

\bibitem{gap2} Mak, K.F., Lee, C., Hone, J., Shan, J. $\&$ Heinz, T.F. Atomically Thin MoS$_2$: A New Direct-Gap Semiconductor. \textit{Phys. Rev. Lett.} \textbf{105}, 136805 (2010).

\bibitem{Gutierrez} Gutiérrez, H.R., \textit{et al.} Extraordinary Room-Temperature Photoluminescence in Triangular
WS$_2$ Monolayers. \textit{Nano Lett.} \textbf{13}, 3447-3454 (2013).

\bibitem{liquid} McCulloch, I., \textit{et al.} Liquid-crystalline semiconducting polymers with high charge-carrier mobility. \textit{Nat. Mater.} \textbf{5}, 328-333 (2006).

\bibitem{impurity} Ghatak, S., Pal A.N. $\&$ Ghosh, A. Nature of Electronic States in Atomically Thin MoS$_2$ Field-Effect Transistors. \textit{ACS Nano} \textbf{5}, 7707-7712 (2011).

\bibitem{hBN} Lee, G.-H., \textit{et al.} Electron tunneling through atomically flat and ultrathin hexagonal boron nitride.\textit{App. Phys. Lett.} \textbf{99}, 243114 (2011).

\bibitem{highMob} Dean, C.R., \textit{et al.} Boron nitride substrates for high-quality graphene electronics. \textit{Nat. Nanotechnol.} \textbf{5}, 722-726 (2010).

\bibitem{shklov} Shklovskii, B.I. $\&$ Efros, A.L. \textit{eds.}, \textit{Electronic Properties of Doped Semiconductors ACS} (Springer Series in Solid State Sciences, Vol. 45, Berlin, 1984).

\bibitem{Efros} Efros, A.L. $\&$ Shklovskii, B.I. \textit{eds.}, \textit{Electron-Electron Interactions in Disordered Systems} (North-Holland, Amsterdam, 1985)

\bibitem{Ramanws3} Ramakrishna, M.H.S.S., \textit{et al.} MoS$_2$ and WS$_2$ Analogues of Graphene. \textit{ Angew. Chem. Int. Edit.} \textbf{49}, 4059-4062 (2010).

\bibitem{Ramanws2} Berkdemir, A., \textit{et al.} Identification of individual and few layers of WS2 using Raman Spectroscopy. \textit{Sci. Rep.} \textbf{3}, 1755 (2013).

\bibitem{SiO2surface} Ishigami, M., Chen, J.H., Cullen, W.G., Fuhrer, M.S. $\&$ Williams, E.D. Atomic Structure of Graphene on SiO$_2$. \textit{Nano Lett.} \textbf{7}, 1643-1648 (2007).

\bibitem{grapheneHyst} Lafkioti, M., \textit{et al.} Graphene on a Hydrophobic Substrate: Doping Reduction and Hysteresis Suppression under Ambient Conditions. \textit{Nano Lett.} \textbf{10},1149-1153 (2010).

\bibitem{Alex} Laiko, E.I., Orlov, A.O., Savchenko, A.K., Il'ichev, E.A. $\&$ Poltoratskii, E.A. Negative magnetoresistance and oscillations of the hopping conductance of a short n-type channel in a GaAs field-effect transistor. \textit{JETP Lett.} \textbf{66}, 1258 (1987).

\bibitem{Das2013} Das, S., Chen, H.-Y., Penumatcha, A.V. $\&$ Appenzeller, J. High performance multilayer MoS2 transistors with scandium
contacts. \textit{Nano Lett.} \textbf{13}, 100 (2013).

\bibitem{Yakimov} Yakimov, A.I., Stepina, N.P. $\&$ Dvurechenskii, A.V. Inelastic resonant tunneling in amorphous silicon microstructures. \textit{Phys. Lett. A} \textbf{194}, 133 (1994).

\bibitem{Savchenko} Savchenko, A.K., \textit{et al.} Resonant tunneling through two impurities in disordered barriers. \textit{Phys. Rev. B} \textbf{52}, R17021 (1995).

\end{thebibliography}
\end{document}